\documentclass[12pt]{article}
\usepackage{amsfonts,bm,bbm}
\usepackage{amssymb}
\usepackage{mathrsfs}
\usepackage{amsmath}	
\usepackage{cite}
\usepackage{graphicx}
\usepackage{rotating}
\usepackage{hyperref}
\usepackage{verbatim}

\csname @addtoreset\endcsname{equation}{section}
\newcommand{\tr}{{\rm Tr}\;}

\newcommand{\beq}{\begin{equation}}
\newcommand{\eeq}{\end{equation}}
\newcommand{\bea}{\begin{eqnarray}}
\newcommand{\eea}{\end{eqnarray}}
\newcommand{\ba}{\begin{array}}
\newcommand{\ea}{\end{array}}
\newcommand{\bit}{\begin{itemize}}
\newcommand{\eit}{\end{itemize}}
\newcommand{\nn}{\nonumber}

\newcommand{\mezzo}{\frac{1}{2}}
\newcommand{\complesso}{{\ \hbox{{\rm I}\kern-.6em\hbox{\bf C}}}}
\newcommand{\reale}{{\hbox{{\rm I}\kern-.2em\hbox{\rm R}}}}
\newcommand{\1}{ \,  \raisebox{+0.14em}{{\hbox{{\rm \scriptsize ]}} \raisebox{-0.2em}{\kern-.8em\hbox{1}}}} \, }  

\newcommand{\p}{\partial}
\newcommand{\var}[2]{\frac{\delta #1}{\delta #2}}

\renewcommand{\a}{\alpha}

\newcommand{\g}{\gamma}

\renewcommand{\d}{\delta}
\newcommand{\D}{\Delta}
\newcommand{\e}{\epsilon}

\renewcommand{\l}{\lambda}

\newcommand{\m}{\mu}

\newcommand{\n}{\nu}
\renewcommand{\r}{\rho}

\newcommand{\vf}{\varphi}

\newcommand{\Across}{\raisebox{-0.25\height}{\includegraphics[width=0.5cm]{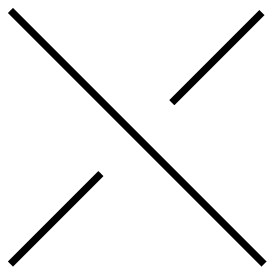}}}
\newcommand{\Bcross}{\raisebox{-0.25\height}{\includegraphics[width=0.5cm]{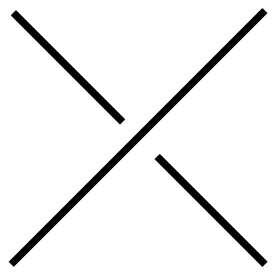}}}
\newcommand{\Asmooth}{\raisebox{-0.25\height}{\includegraphics[width=0.5cm]{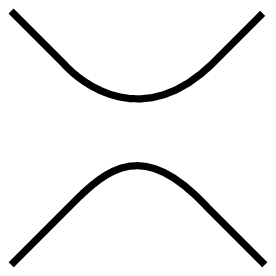}}}
\newcommand{\Bsmooth}{\raisebox{-0.25\height}{\includegraphics[width=0.5cm]{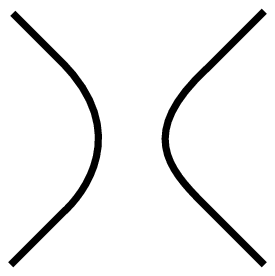}}}
\newcommand{\Rcurl}{\raisebox{-0.25\height}{\includegraphics[width=0.5cm]{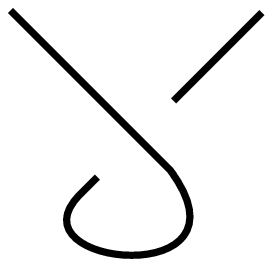}}}
\newcommand{\Lcurl}{\raisebox{-0.25\height}{\includegraphics[width=0.5cm]{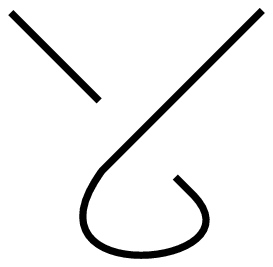}}}
\newcommand{\Arc}{\raisebox{-0.25\height}{\includegraphics[width=0.5cm]{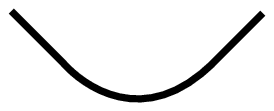}}}
\newcommand{\duecross}{\raisebox{-0.31\height}{\includegraphics[width=0.30cm]{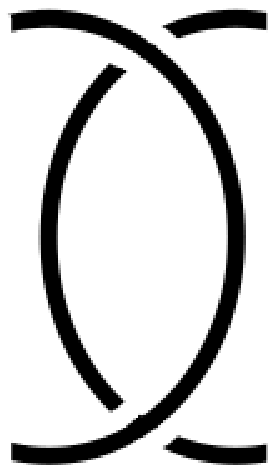}}}

\begin{document}

\begin{titlepage}
\begin{flushright}
hep-th/??????\\
CECS-PHY-10/08
\end{flushright}
\vspace{0.8cm}
\begin{center}
\renewcommand{\thefootnote}{\fnsymbol{footnote}}
{\Large \bf Kauffman Knot Invariant from}
\vskip 5mm
{\Large \bf SO(N) or Sp(N) Chern-Simons theory}
\vskip 5mm
{\Large \bf and the Potts Model}
\vskip 25mm
{\large {Marco Astorino\footnote{marco.astorino@gmail.com}}}\\
\renewcommand{\thefootnote}{\arabic{footnote}}
\setcounter{footnote}{0}
\vskip 10mm
{\small \textit{
Instituto de F\'{\i}sica,\\ 
Pontificia Universidad Cat\'olica de Valpara\'{\i}so \\
and\\
Centro de Estudios Cient\'{\i}ficos (CECS), Valdivia,\\ 
Chile\\}
}
\end{center}
\vspace{2.0cm}
\begin{center}
{\bf Abstract}
\end{center}
{\small The expectation value of Wilson loop operators in three-dimensional SO(N) Chern-Simons gauge theory gives a known knot invariant: the Kauffman polynomial. Here this result is derived, at the first order, via a simple variational method.
With the same procedure the skein relation for Sp(N) are also obtained. Jones polynomial arises as special cases: Sp(2), SO(-2) and SL(2,$\mathbb{R}$).}\\
{\small These results are confirmed and extended up to the second order, by means of perturbation theory, which moreover let us establish a duality relation between SO($\pm$N) and Sp($\mp$N) invariants.}\\ 
{\small A correspondence between the firsts orders in perturbation theory of SO(-2), Sp(2) or SU(2) Chern-Simons quantum holonomies and the partition function of the $Q=4$ Potts Model is built.}
\end{titlepage}

\section{Introduction}

In a milestone work \cite{witten} Witten realised that the expectation value of a Wilson loop, computed with a three-dimensional Chern-Simons action measure, was a knot invariant. This is due to the fact that the Wilson loops are observables for Chern-Simons theories, having therefore diffeomorphism invariant expectation values. More in general this feature stems from the property that such a quantum field theory manifests general covariance, which in turn is a consequence of the metric independent structure: any physical quantity computed in this framework is a topological invariant.  \\
In practise, for SU(N) Chern-Simons field theory, the resulting knot invariant is the HOMFLY polynomial, which in particular specialises into the Jones polynomial in the case of SU(2). These outcomes were derived through both conformal field theory (as in \cite{witten}) or perturbative quantum field theory (see for instance \cite{guadagnini}). 
But a simpler heuristic derivation was proposed in \cite{cotta} and \cite{smolin} (for reviews see also \cite{kauffman} and \cite{ga-pu}), at least up to the first order in the inverse coupling constant of the theory. It is based on a variational approach: it studies the behaviour in the expectation value of the Wilson loop when one performs small geometric deformation.\\
In the conformal field theory scheme similar results have been found in \cite{kcp}, \cite{horne} and \cite{wu} for several other groups: SO(N), Sp(N), SU(n$|$m) and OSp(m$|$2n).\\
It would be interesting to test whether the variational procedure, which is expressly realised to reproduce the HOMFLY polynomial from SU(N) gauge theory, may apply also in different contexts. In section \ref{variational-approach} are studied the SO(N), SL(N,$\mathbb{R}$) and Sp(N) cases.\\
The results obtained are moreover analysed in section \ref{perturbative-approach} by means of the more rigorous standard perturbation theory and extended  up to the subsequent order, the second.\\ 
Finally in section \ref{correspondence} we try to interpret these results from the statistical mechanic point of view, trying to connect the holonomies's first order expansion to one of the more famous lattice statistical system: the Q-Potts Model\footnote{We are referring to the standard two dimensional Potts model, not to some variant with multiple Boltzmann weights, which in much literature are misleading called in same way.}; which at the moment remains unsolved apart for its easiest personification when Q=2, the Ising model.
We start (section \ref{chern-kauffman}) introducing the notation and summarising the fundamental properties of Chern-Simons theory and Kauffman polynomial that are useful in derivation of skein relations.\\

\section{Chern-Simons theory and Kauffman polynomial}
\label{chern-kauffman}

Let's consider a Chern-Simons theory for a gauge field connection one-form $A= A^a_{\ \m} (x) T^a dx^\m$ valued in a generic semi-simple Lie algebra $\mathfrak{g}$, with action: 
\begin{equation}   
\mathcal{L_{CS}}[A] 
                    = \frac{k}{4 \pi} \int_{\mathcal{M}^3} d^3 x \ \frac{\e^{\m \n \r}}{2} \ \left(A^a_{\ \m} \p_\n  A^a_{\ \r} - \frac{1}{3} A^a_{\ \m} A^b_{\ \n} A^c_{\ \r} f^{a b c}  \right)    \nn
\end{equation}     
where $\mathcal{M}^3$ is a compact three-dimensional manifold whose coordinates are labelled by Greek letters ($\m, \n, \r,...$); while the internal group indices will be denoted by Latin letters ($a, b ,c,...$). The Lie algebra is spanned by generators $T^a, T^b,\dots$, obeying the commutation relations $[T^a,T^b]= i f^{abc} T^c$ and normalised as follows: $\tr(T^a T^b) = \mezzo \d^{a b}$. 

This action got several notable properties: $(i)$ it changes by $2 \pi k n_g$ under a gauge transformation $A_\m \rightsquigarrow A'_\m=g^{-1}A_\m g -i g^{-1}(\p_\m g)$ ($n_g$ is the degree of the mapping $g:\mathcal{M}^3\rightarrow\mathcal{G}$); thus, $ \forall \ k \in \mathbb{Z} $, $\mathrm{exp}(i \mathcal{L_{CS}})$ is a complete gauge invariant quantity that will play the r\^ole of the path integral measure. $(ii)$  The curvature of the gauge field at the point $x \in \mathcal{M}^3$ is given by:
$$ F^a_{\ \m\n}(x)= \frac{4 \pi }{k} \e_{\m\n\l} \ \var{\mathcal{L_{CS}}[A(x)]}{A^a_{\ \l}(x)} $$

We will interested in computing expectation values $\langle W(\g) \rangle$ for Wilson loops $W_\g[A]$ along closed paths $\g$, that in fact may be thought as a knot on $\mathcal{M}^3$, defined as follows:
\bea
     W_\g[A] &=& \tr \Big[\mathrm{P} \ \mathrm{exp}\Big(i \oint_\g A_\m dx^\m \Big) \Big] \nn \\
     \langle W(\g) \rangle &=& \mathcal{Z}^{-1} \ \int \mathscr{D}A \ \mathrm{exp}\left( i \ \mathcal{L_{CS}}[A]\right) \ W_\g[A] \nn
\eea
In this notation $\g$ represents both common knots $\g(t):I\rightarrow\mathcal{M}^3$ and n-component knots, also called knot-links, $\g(t_1,t_2,\dots,t_n)=(\g_1(t_1),\g_2(t_2),\dots,\g_n(t_n)):I_1 \times I_2 \times \dots I_n \rightarrow \mathcal{M}^3$.  In the latter case $\langle W(\g) \rangle =  \langle W(\g_1) W(\g_2) \dots W(\g_n) \rangle$. Without losing generality one may think the compact interval $I_i =[0,1]$  and  $\g(0)=\g(1)$ in order to have closed paths.
The fact that Chern-Simons action is independent of the particular choice of a metric on the three-manifold suggests that the Wilson loop expectation values may capture some invariant or topological characteristic 
of the system's geometry: either that of the knots or of the manifold itself.\\

\begin{figure}[h]
\begin{center}
\includegraphics[angle=0, scale=0.6] {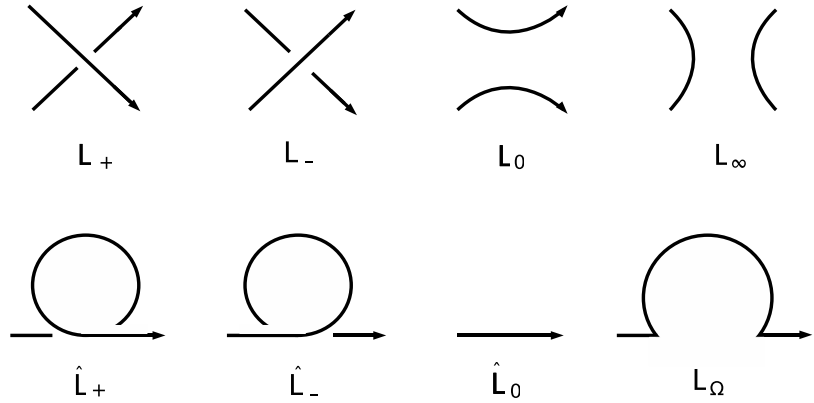}
\end{center}
\caption{\small{Different crossing configurations involved in the skein relations.
Dealing with unoriented links, arrows can be ignored because they carry no sensitive information.}}
\label{crossing}
\end{figure}

Now we introduce the Kauffman polynomial which is a regular isotopy invariant of knots and, if suitably normalised, becomes an ambient isotopy invariant. Actually we will deal with its equivalent Dubrovnik version. To each knot-link there is associated a finite Laurent polynomial $D_K=D_K(a,z)$ of two variables with integer coefficients, such that if $K_1\sim K_2$, then $D_{K_1} = D_{K_2}$ (while the reverse is not necessary true). The polynomial can be constructed, as in \cite{kauffman2} or \cite{wu-knot}, by the following rules\footnote{Sometimes, as in \cite{kauffman}, can be found a different normalisation for $D_K$: $iii)' \ \ D(\bigcirc) = 1 + \frac{a-a^{-1}}{z}$; in our notation $1 + \frac{a-a^{-1}}{z}$ will result the $\langle \bigcirc \rangle$ 's normalisation.} (see figure \ref{crossing} for notation, $\bigcirc$ stands for the unknotted circle):
\bea
\label{skein}
      &i)&    \qquad       D(L_+)-D(L_-) =  z \ [ D(L_0) - D(L_{\infty})] \nn  \\ 
      &ii)&   \qquad       D(\hat{L}_{\pm}) = a^{\pm} D(\hat{L}_0)     \\ 
      &iii)&  \qquad       D(\bigcirc) = 1 \nn
\eea
In $i)$ and $ii)$ the small diagrams $\{L_k\}_{k=\pm,0,\infty}$ stand for larger link diagrams that differ only as indicated by the smaller ones. 
Starting from any knot-links K and using recursively Reidemeister moves and the skein relations (\ref{skein}) at each diagram's crossing, one can obtain uniquely its regular isotopy invariant $D_K(a,z)$. It is possible to normalise $D_K$ by a factor that take into account also eventual contributions of twists. For this purpose is used the {\it writhe} $w(K)=\sum_p \e(p)$, where $p$ runs over all crossing in $K$ and $\e(L_{\pm})=\pm1$ is the sign of the type of crossing. So finally we are able to define a genuine ambient isotopy invariant: the normalised Kauffman-Dubrovnik polynomial\footnote{While $D_k$ is defined for unoriented knots, to calculate the writhe in $Y_K$ one needs to define an orientation. At the end the orientation does not affect the result for knots but it affects the invariant polynomial in case of proper links. Thus $Y_K$ is said to be defined for semi-oriented knot-links.}:
\beq
     Y_K(a,z)=(a)^{-w(K)} D_K(a,z) \nn
\eeq

\section{Variational derivation of the skein relation}
\label{variational-approach}

It's well known (see \cite{kauffman} for details) that the Wilson loops satisfy the following differential equations:
\bea
    \d_A  W_\g[A] &=& \frac{\d W_\g[A]  }{\d A^a_{\ \m}(x)} = i \ T^a \ dx^\m \ W_\g[A]  \nn \\
    \d_{\g_x} W_\g[A] &=& i F^a_{\ \m\n} T^a \ dx^\m dx^\n W_\g[A] \nn
\eea
where $\d_{\g_x}$ is the variation corresponding to an infinitesimal deformation of the loop $\g$ in the neighbourhoods of a point $x$.
It's then possible to compute this variation for an expectation value of a Wilson line along a knotted path $\g$ and to use it to obtain a formula for the switching identity $ \langle W(\hat{L}_+) \rangle -\langle W(\hat{L}_-) \rangle $ as\footnote{Proposition 17.4 and theorem 17.5 of \cite{kauffman}.} follows:
\beq 
\label{varw0}
     \d_{\g_{x}} \langle W(\g) \rangle = - \frac{4 \pi i}{k} \frac{1}{\mathcal{Z}} \int \mathscr{D}A \ \mathrm{exp} \ \left(i \ \mathcal{L_{CS}}[A]\right) \ \left[  \e_{\m\n\l} dx^\m dx^\n dy^\l \right] [ \sum_a T^a T^a ] W_\g[A]  
\eeq
Note that studying the formal properties of this integral three assumptions are always used: $i)$ the limits of differentiation and integration commute: $\d_{\g_x} \langle W_\g[A] \rangle = \langle \d_{\g_x} W_\g[A] \rangle$; $ii)$ integrating by parts it's possible to discard the boundary term; $iii)$ the existence of an appropriate functional measure on this moduli space. \\
From the previous equation one is able to write the switching identity $ \langle W(\hat{L}_+) \rangle -\langle W(\hat{L}_-) \rangle $. The quantity $\left[  \e_{\m\n\l} dx^\m dx^\n dy^\l \right]$ is dimensionless and, whether properly normalised, can be thought -1,0 or 1. Then (\ref{varw0}) has a standard interpretation (we follow \cite{kauffman}) if one calls the operator, which in some sense enclose the loop's small deformation, $C= \sum_a T^a T^a $:
\beq
\label{varw}
               \langle W(\hat{L}_+) \rangle -\langle W(\hat{L}_-) \rangle = - \frac{4 \pi i}{k} \langle C \ W(\g) \rangle 
\eeq 
Graphically $\langle C \ W(\g) \rangle$ is represented in the l.h.s of figure's \ref{tt} equation.
Note that the sign is a convention which may be reversed exchanging $\hat{L}_+ \leftrightarrow \hat{L}_-$.\\
Till this point the whole model has been valid for a generic gauge group $\mathcal{G}$. In particular was successfully used in the literature to reproduce the Witten's result for HOMFLY polynomials from the SU(N) group. Instead in this paper we specialise our study to two particular algebras which have simple Fierz identities: the ones associated to the orthogonal group SO(N) and the symplectic group Sp(N), for a generic N.

\subsection{SO(N) and Kauffman polynomial}
\label{son-sect}

\begin{figure}[ht]
\begin{center}
\includegraphics[angle=0, scale=0.6] {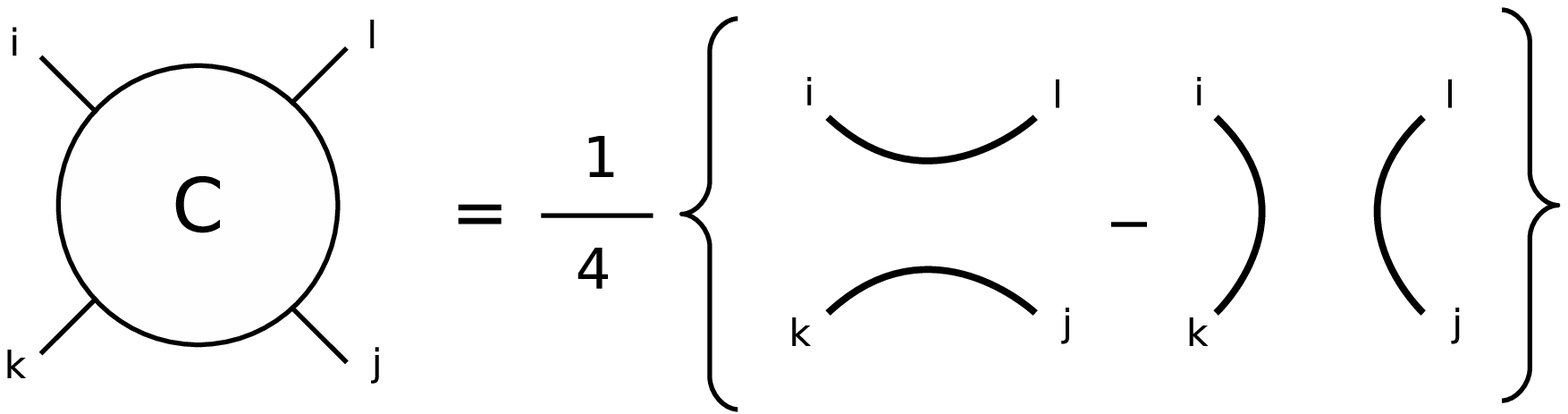}
\end{center}
\caption{\small{Abstract diagrammatic representation of Fierz identity for SO(N)}}
\label{tt}
\end{figure}

Here the features of the algebra under consideration begin to play an important r\^ole. In fact to evaluate the operator $C$ one needs to use the Fierz identity; in particular we have for SO(N) in the fundamental representation (in \cite{cvitanovic} Fierz identities are presented for almost all semi-simple Lie groups):
\beq
      \sum_a (T^a)^i_{\ j} (T^a)^k_{\ l} = \frac{1}{4} \left( \d^i_{\ l} \d^k_{\ j} - \d^{ik} \d_{jl} \right)  \nn
\eeq
This expression in the Baxter's abstract tensor notation (see \cite{kauffman}) reads as the diagrammatic relation drawn in figure \ref{tt}.
Hence, substituting in (\ref{varw}) the Fierz identity we have:
\beq \label{skeinso}
     \langle W(L_+) \rangle -\langle W(L_-) \rangle = - \frac{\pi i}{k} \big[ \langle W(L_0) \rangle -\langle W(L_{\infty}) \rangle  \big]
\eeq
To get in touch with the known results, one has to take the limit of $k >> 1$, namely the analogous of the first order perturbation expansion, thus the previous expression reduces to:
\beq  
    \langle W(L_+) \rangle -\langle W(L_-) \rangle
                                                    = \big(q -q^{-1} \big)  \ \big[ \langle W(L_0) \rangle -\langle W(L_{\infty}) \rangle  \big] \nn
\eeq
These are exactly the skein relations that are found by means of the original Witten's method based on conformal field theory arguments (see \cite{kcp} and \cite{horne}), once $q:=\mathrm{exp}(-\frac{\pi i}{2k})$ is defined\footnote{\cite{horne} uses a different killing metric normalisation for the Lie algebra generators; in order to compare with it one has to define a slightly different $q:=\textrm{exp}(-\frac{\pi i}{k})$. \cite{kcp} uses an inverse definition of the writhe and of the crossing diagrams, so what they call $\alpha = a^{-1}$ and their $q$ is our $q^{-1}$.}.
So is not difficult to see that $D_K= \langle W(K) \rangle / \langle W(\bigcirc) \rangle$ fulfils the definition of Dubrovnik polynomial (normalised as in \cite{kauffman2} and \cite{wu-knot}\footnote{Clearly if {\it write}-normalised by a factor $a^{-w(K)}$ (where $w(L_\pm)=\pm 1$) $D_K(a,z)$ became an ambient isotopy invariant.}), with $z=(q-q^{-1})$. The only thing that remains to fix is the value of $a$ such that $\langle W(\hat{L}_+) \rangle=  a \langle W(\hat{L}_0) \rangle$. This can be done considering the closure of the path in the skein relation (\ref{skeinso}), as shown in the figure below:
\begin{figure}[ht]
\caption{\small{Diagrammatic closure of the SO(N) skein relation (\ref{skeinso})}}
\begin{center}
\includegraphics[angle=0, scale=0.7] {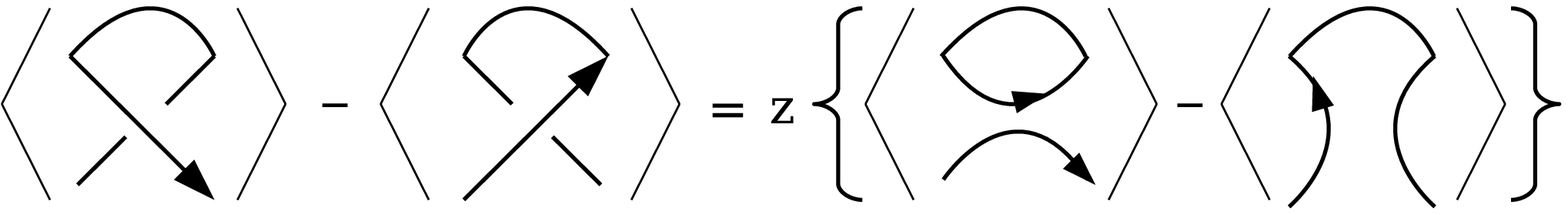}
\end{center}
\label{closure}
\end{figure}
\bea \label{norm}
      \langle W(\hat{L}_+) \rangle - \langle W(\hat{L}_-) \rangle &=& - \frac{\pi i}{k} \big[ \langle W(\bigcirc \ \hat{L}_0) \rangle - \langle W(\hat{L}_0) \rangle \big] \nn \\
     a \langle W(\hat{L}_0) \rangle - a^{-1} \langle W(\hat{L}_0) \rangle &=& - \frac{\pi i}{k} \big[ (N-1) \langle W(\hat{L}_0) \rangle \big]   
\eea
Solutions for (\ref{norm}) are $a=q^{N-1}$ or $a=-q^{1-N}$, which however gives rise at an equivalent $D_K$ polynomials\footnote{Just redefine $q\rightarrow \tilde{q}=-q^{-1}$ to verify the second root branch redundancy.}. The factor $N$ comes from the diagrammatic tensor interpretation of the unknot circle, that is $\d_i^{\ i}=N$. It's worth to observe that these Dubrovnik-Kauffman polynomials $D_K(a= -q^{1-N},z=q-q^{-1})$ do not run out all the original ones, but constitute a smaller subset depending on the fact that $a$ assumes only discrete values depending on $N$ (which generally is thought in $\mathbb{N}$). \\
The consistency check up to the $1/k$ order proposed in \cite{cotta} is intrinsically satisfied using the quadratic Casimir operator of $\mathfrak{so}(N): \mathbbm{1} (N-1)/4$. Moreover the variational first order approach, can be generalised to subsequents orders with the same arguments presented in \cite{bruegmann} and \cite{gambini} for SU(N) groups. But we will prefer explore the subsequent order of the expansion (see section \ref{perturbative-approach}) through a different method based on the standard quantum field theory of perturbations.\\
Finally note that the original Jones polynomial  $a^{-w(K)} D_K(\bar{a}= -q^{3},\bar{z}=q-q^{-1})$ is not included in this sub-class of Kauffman polynomial, unless choosing unconventionally $N=-2$ (once the polynomial is analytic continued for all integers values of N).\\ 
Negative dimensions group theory is a powerful technique, first introduced by Penrose, to calculate algebraic invariants (see \cite{cvitanovic-libro}, \cite{maru} and \cite{parisi}). In that case it relates the Casimirs and Young tableau of SO(-2) to the ones of Sp(2). Some speculation about this possibility are done in the next subsection, while a more rigorous treatment is done on section \ref{perturbative-approach}. \\
One may be puzzled not to come across Jones polynomial for the SO(3) group which is locally isomorphic to SU(2) where this relation holds. The reason for this mismatch is based on the fact that in this context, more than groups similarities, the Lie algebras invariants play a key r\^ole. \\
Actually, as also for SL(2,$\mathbb{R}$) generators the same SU(2) Fierz identity for the $C$ operator holds
, Jones polynomial can be recovered with the same procedure of \cite{cotta}. It is not surprising because $\mathfrak{sl}(2,\mathbb{R})$ is the real split form of the $A_1$ algebra (known also as the $\mathfrak{sl}(2,\mathbb{C})$ algebra by an abuse of notation), while $\mathfrak{su}(2)$ is the real compact one. 

\subsection{Sp(N) skein relations and Jones Polynomial for Sp(2)}
\label{varspn}

\begin{figure}[ht]
\begin{center}
\includegraphics[angle=0, scale=0.6] {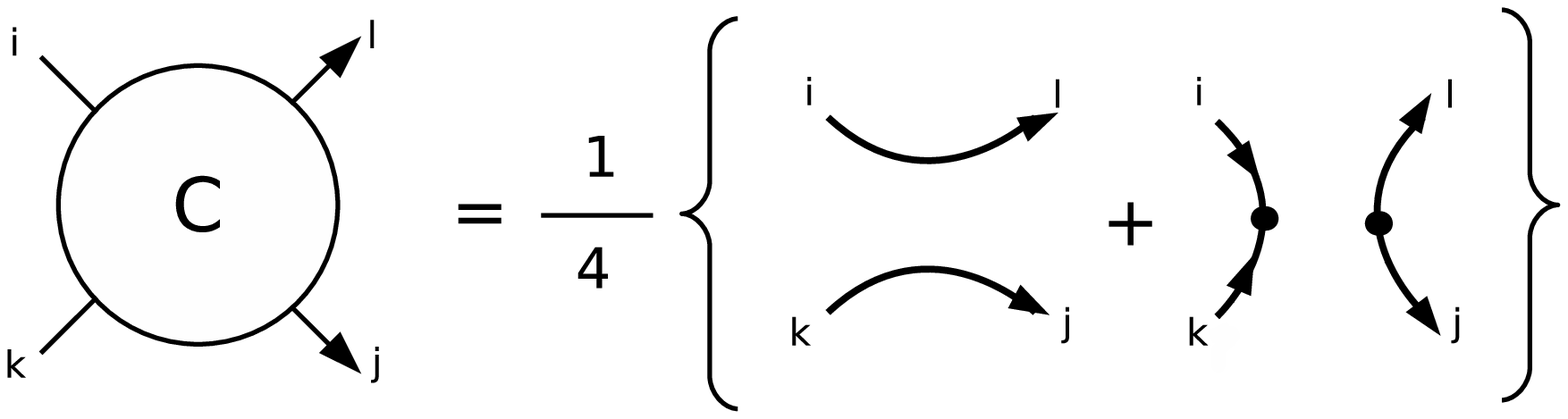}
\end{center}
\caption{\small{Fierz identity for Sp(N), dots represent points where orientations of the line change.}}
\label{tt-sp}
\end{figure}

In this section we consider the Symplectic group Sp(N), for even N; apart from the relation with SO(-N) it is an interesting case for itself. Its Fierz identity (see again \cite{cvitanovic}) for the generators in the fundamental representation is:
\beq
      \sum_a (T_a)^i_{\ j} (T_a)^k_{\ l} = \frac{1}{4} \left( \d^i_{\ l} \d^k_{\ j} + f^{ik} f_{jl} \right)  \nn
\eeq
where $f^{ij}=-f^{ji} \ , \ f^{ij}f_{jk}=\d^i_{\ k}$. As the fundamental representation of this group is pseudoreal, unlike SO(N), the orientation should not be neglected as it is shown in figure \ref{tt-sp}.\footnote{In \cite{horne} another approach (which has the advantage that leaves the Wilson lines unoriented) is also presented, but not preferred as requires the specific choice of a ''time'' direction, which breaks the topological invariance because it is no longer possible to freely rotate the Wilson lines.} 
Plugging this Fierz identity for Sp(N) into eq. (\ref{varw}) one fits the same skein relation of \cite{horne} which is obtained by a totally different approach.\footnote{We refer to the one drawn in figure 17 of \cite{horne}} 

\begin{figure}[h]
\begin{center}
\includegraphics[angle=0, scale=0.75] {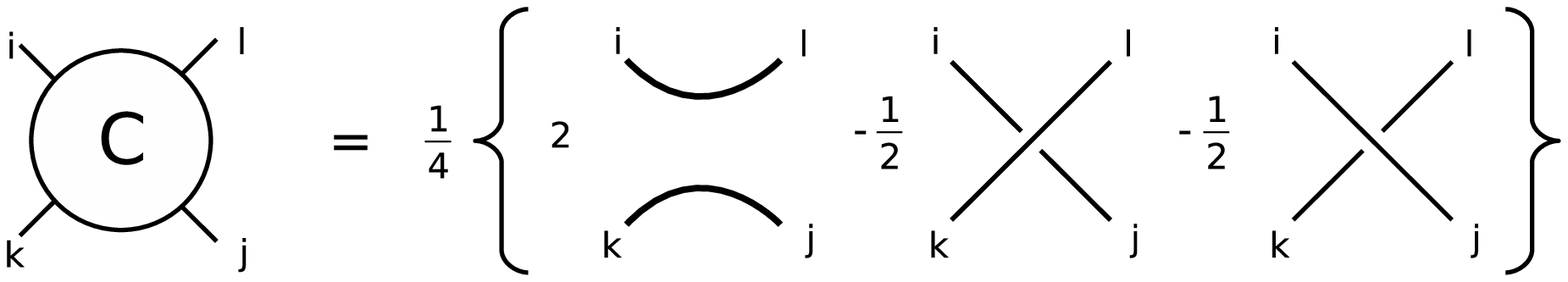}
\end{center}
\caption{\small{Diagrammatic representation of Fierz identity for Sp(2)}}
\label{tt-sp2}
\end{figure}

There is a particular case where those computation are easily\footnote{Even without the oriented diagram notation which is unnecessary heavy for Sp(2). One might work, in a complete compatible way, with the arrowed diagrams but paying the price of redefining appropriate oriented Reidemeister moves and oriented Kauffman state bracket as described in cap $6^0$ of \cite{kauffman} and \cite{horne}.} carried on till get its knot invariant: N=2, just the one suspected to be related to the Jones polynomial, as we saw in section \ref{son-sect}. In fact for Sp(2) the antisymmetric matrix $f^{ij}$ may be straight interpreted, without losing generality, as the Levi-Civita tensor $\e^{ij}$ and its inverse $f_{ij}=-\e_{ij}$
. Hence the algebraic (eq. (\ref{fierzsp2al})) and diagrammatic (fig. \ref{tt-sp2}) representations of the C operator appear respectively as follows:
\bea \label{fierzsp2al}
    \sum_a (T_a)^i_{\ j} (T_a)^k_{\ l} = \frac{1}{4} \left( \d^i_{\ l} \d^k_{\ j} - \e^{ik} \e_{jl} \right) 
                                           = \frac{1}{4} \left( 2 \d^i_{\ l} \d^k_{\ j} - \d^i_{\ j} \d^k_{\ l} \right)  
\eea

Now substituting the Fierz identity for Sp(2) into (\ref{varw}) we have:
\bea
    \langle W(L_+) \rangle - \langle W(L_-) \rangle &=& - \frac{2 \pi i}{k} \langle W(L_0) \rangle + \frac{\pi i}{2k} \langle W(L_+) \rangle + \frac{\pi i}{2k} \langle W(L_-) \rangle  \nn \\
    \left(1- \frac{\pi i}{2k} \right) \langle W(L_+) \rangle &-& \left(1 + \frac{\pi i}{2k} \right) \langle W(L_-) = - \frac{2 \pi i}{k} \langle W(L_0) \rangle \nn \\
     q \langle W(L_+) \rangle - q^{-1} \langle W(L_-) \rangle &=& \tilde{z} \langle W(L_0) \rangle \nn 
\eea
Where $q$ is the same of section \ref{son-sect}, while it is defined $\tilde{z}:= -\frac{2 \pi i}{k} = x -x^{-1}$ if we call $x:= \textrm{exp}(-\frac{\pi i}{k})$. Again we are considering at this stage $k>>1$, i.e these equalities hold up to first order in the inverse coupling constant of the theory\footnote{The first order consistency check proposed in \cite{cotta} is trivially satisfied using, this time, the quadratic Casimir operator of $\mathfrak{sp}(2) : \ 3\mathbbm{1}/4$}.
Closing the path in the previous skein relation as done for SO(N) we will be able to get a constraint that reduces one variable dependence:
\bea
     q \langle W(\hat{L}_+) \rangle - q^{-1} \langle W(\hat{L}_-) \rangle &=& \tilde{z} \langle W(\hat{L}_0 \ \bigcirc) \rangle \nn \\
     a q \langle W(\hat{L}_0) \rangle - a^{-1} q^{-1} \langle W(\hat{L}_0) \rangle &=&  x^2-x^{-2} \langle W(\hat{L}_0) \rangle \nn \\
                                \Longrightarrow  \qquad a q &=& x^2   \nn
\eea
As before the second root $a q =-x^{-2}$ leads exactly to the same results. So at large values of $k$ for a normalised (to be a) expectation value $ P(K) = a^{-w(K)} \langle W(K) \rangle/\langle W(\bigcirc) \rangle  $ the original one variable Jones polynomial follows directly:
$$ x^2 P(L_+) -x^{-2} P(L_-) = (x - x^{-1}) P(L_0)  $$ 
So actually the estimation suggested by negative dimension group theory seems to work reliably. As it's here proved the Sp(2) Chern-Simons expectation values of a Wilson knot-link gives the Jones polynomial invariant for the same link.

\section{Perturbative Quantum Field approach}
\label{perturbative-approach}

It's worth analysing the heuristic previous section's results in a more carefully way. We opt for the standard quantum field theory of perturbation as developed for the SU(N) group in \cite{guadagnini}, which maybe got the disadvantage of being less qualitative from a geometrical point of view but got the benefit of being more analytically quantitative. The fact of being, in principle, a different approach also adds some guaranties on the consistency  of the check. Not least this method let us push the expansion, in the inverse coupling constant $k$, to one order further.\\
Note that for this procedure a framing of the knot is needed; in this paper is always used the \emph{vertical frame} defined as the one that got linking number equal to the writhe of the knot $\varphi_f(K) = w(K)$. Framed knots can be thought as bands, so in this picture a writhe for a knot represents a band twist. As Kauffman polynomial are regular isotopy invariant, twisted bands are the most suitable objects to be described with.
The expectation value for the Wilson loop computed along a vertical framed, m-component ($C_1,C_2, ... , C_m$) knot-link $K$ in a Chern-Simons theory for a generic semisimple group $\mathcal{G}$ is given at second order by: 
\bea
\label{w2ord}
\langle W (K) \rangle &=&  \Big(\prod_{k=1}^{m} \textrm{dim} \ T_k \Big) \Big\{1 - i \Big(\frac{2 \pi}{k} \Big) \sum_{k=1}^{m} c_2(T_k) \varphi_f (C_k) \\                    &-&  \Big(\frac{2 \pi}{k} \Big)^2 \ \sum_{k=i}^{m} \Big[ \mezzo c_2^2(T_k) \varphi^2_f(C_k) - c_v c_2(T_k) \rho(C_k) \Big] \nn \\
                      &-&   \Big(\frac{2 \pi}{k} \Big)^2 \sum_{k \neq \ell} c_2(T_k) c_2(T_\ell) \Big[ \vf_f(C_k) \vf_f(C_\ell) + \frac{\chi^2(C_k,C_\ell)}{\textrm{dim} \ \mathcal{G}} \Big] + O \Big( \frac{1}{k^3} \Big) \Big\} \nn
\eea
where $T$ stands for the fundamental representation, $\chi(C_k,C_\ell)$ is the Gauss linking number between the two curves $C_k$ and $C_\ell$, $\big(c_2(T)\big)_{i}^{\ j}=\sum_a (T^a)_i^{\ k} \  (T^a)_k^{\ j}$ is the quadratic Casimir in the fundamental representation, $c_v$ the quadratic Casimir in the adjoint representation, $\rho(C)$ is an ambient isotopy invariant characteristic of the knot under consideration. $\r(C)$ represents the second coefficient of the Alexander-Conway polynomial and is related with Arf- and Casson-invariants; in practise it is not easy to compute apart from small knots.\\
Our aim is now, with the help of (\ref{w2ord}), to find the value of $a$ appearing in (\ref{skein}-$ii$) in terms of its expansion in $(2 \pi/k)$. The effect of changing the frame of a link component $C_i$ by $\D \varphi_f(C_i)= \D w (C_i)= \pm 1$ (or adding a twist in the band picture) reflects in the entire Wilson loop expectantion value by:
$$ \langle W(K_{\varphi \pm 1}) \rangle =  \a^{(\pm)} \langle W(K_{\varphi}) \rangle $$
\beq \label{alpha}  \a^{(\pm)} =  1 \mp i \Big( \frac{2 \pi}{k} \Big) c_2(T) - \mezzo \Big( \frac{2 \pi}{k} \Big)^2  c_2^2(T) +  O \Big( \frac{1}{k^3} \Big)  \eeq
So we find $a^{\pm1}=\a^{(\pm)}$, taking into account $D_K=\langle W (K) \rangle / \langle W (\bigcirc) \rangle$ as previously defined on section \ref{son-sect}. While (\ref{skein}-$iii$) is trivially satisfied, is possible to extract the value of $z$ from (\ref{skein}-$i$), for instance applying it to the Hopf-link $\mathcal{HL}$.

\begin{figure}[h]
\begin{center}
\includegraphics[angle=0, scale=0.85] {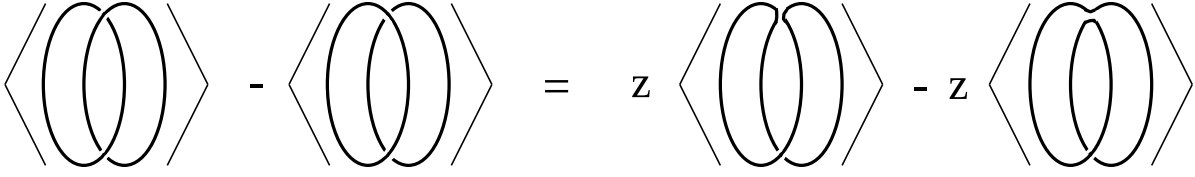}
\end{center}
\caption{\small{Skein relation \ref{skein}-$i)$ applied to the the upper $\mathcal{HL}$ crossing}}
\end{figure}
That is closing the skein relation (\ref{skein})-$i$ as shown above one gets the following expression:
$$  D_{\mathcal{HL}}-D_{\bigcirc \bigcirc} = z  (a -a^{-1} ) D_{\bigcirc}  $$
written in term of relatively easy objects that can be computed directly from (\ref{w2ord}), using as in \cite{guadagnini}, $\r(\bigcirc)=-1/12$:
\bea \label{samples}
     D_{\bigcirc \bigcirc} &=&    N \left[ 1 - \frac{1}{12} \left( \frac{2 \pi}{k} \right)^2 c_v c_2(T) + O \left( \frac{1}{k^3} \right) \right]                     \\ 
     D_{\mathcal{HL}} &=& N \left[ 1 - \frac{1}{12} \left( \frac{2 \pi}{k} \right)^2 c_v c_2(T) - \left( \frac{2 \pi}{k} \right)^2 c_2^2(T) \frac{2}{\textrm{dim} \mathcal{G}} + O \left( \frac{1}{k^3} \right) \right] \nn
\eea
An alternative way to find $z$ is imposing the equality between Kauffman $D_K(a,z)$ polynomials obtained from the skein relations (\ref{skein}) with the expansion of $\langle W (K) \rangle / \langle W (\bigcirc) \rangle$ coming from (\ref{w2ord}). But this could be done just for the few simple knots where $\rho(K)$ can be calculated, so may be here regarded as a self-consistency check.\\
That's the point where the algebraic properties of the gauge groups come out; for the groups we are interested in, they are summarised in the following table:
\begin{center}
\begin{tabular}{|c|cccc|}
	\hline
  &  dim $\mathcal{G}$  &  dim $T$  &  $c_2$  & $c_v$   \\
	\hline
SO(N)   & $N(N-1)/2$ & $N$ & $(N-1)/4$ & $(N-2)/2$   \\
Sp(N)   &  $N(N+1)/2$ & $N$ & $ (N+1)/4$ & $(N+2)/2$   \\
SU(N)   & $N^2-1$  & $N$ & $(N^2-1)/2N$  &  $N$  \\
	\hline
\end{tabular}
\end{center}
hence, from (\ref{alpha}), we get respectively for SO(N) and Sp(N) the following values for $a$
\bea \label{aa}
  a_{SO(N)} &=&  1 - i \left( \frac{2 \pi}{k} \right) \frac{N-1}{4} - \mezzo \left( \frac{2 \pi}{k} \right)^2  \left( \frac{N-1}{4} \right)^2 +  O \left( \frac{1}{k^3} \right)  \\
  a_{Sp(N)} &=&  1 - i \left( \frac{2 \pi}{k} \right) \frac{N+1}{4} - \mezzo \Big( \frac{2 \pi}{k} \Big)^2  \left( \frac{N+1}{4} \right)^2 +  O \left( \frac{1}{k^3} \right)  \nn
\eea
while for both orthogonal and symplectic groups the value found for z is:
\beq \label{z} z = - \frac{ i \pi}{k} + O \Big( \frac{1}{k^3}   \Big)  \eeq
These results are consistent with the ones found in the previous section by means of the variational method both for SO(N) and Sp(2). Moreover (\ref{aa}) and (\ref{z}) extend the series expansion in $2 \pi/k$ up the second order. The fact that $z$ has not the quadratic contribution could be guessed from the very beginning because of the peculiar property of the Chern-Simons Lagrangian: the inversion symmetry. This implies that a change in the sign of the coupling constant $k$ is compensated by the inversion of the orientating of the manifold. When a knot $K$ is embedded in $\mathcal{M}^3$ the change of orientation of the manifold corresponds to a $\pi$ rotation or its mirror image $\tilde{K}$, so $\langle W(K)\rangle \big|_k=\langle W(\tilde{K}) \rangle \big|_{-k} $. On the other hand from skein relations (\ref{skein}) is easy to see that $ D_K(a,z) = D_{\tilde{K}}(a^{-1},-z) $; combining it with the inversion symmetry one gets some restriction on the k-functional dependence of the variables $a$ and $z$: 
\beq    \label{restriction}    a(k)=a^{-1}(-k) \qquad z(k)=-z(-k)    \eeq 
So even powers of $k$ were not expected in the $z$ expansion; as one can see (\ref{aa}) and (\ref{z}) fulfil the constraints (\ref{restriction}). 
The easiest functions that are compatible with the series expansions (\ref{aa})-(\ref{z}), their restrictions (\ref{restriction}) and the samples (\ref{samples}) are:
$$ a = \textrm{exp} \left[ -i \frac{2\pi}{k} c_2(T)  \right] \qquad z=-2 \ i \ \textrm{sin}\left( \frac{\pi}{2k} \right) $$ \\
Furthermore observe that in the groups table there is a value of N for whom two lines match: for $N=2$ all the values for Sp(2) and SU(2) coincide. So the expectation value of a Wilson loop along a generic knot K agrees in both cases. This special point is the one where the HOMFLY and Kauffman polynomials overlap to give the Jones polynomial. This is exactly the same result we have found with the variational approach in section \ref{varspn}, but now extended to the second order. Another interesting feature that can be read from the table is the analogy between the quantities of SO(-N) and Sp(N), in particular one can note in (\ref{w2ord}) as Wilson loop expectation values of a SO(-N)-Chern-Simons theory for a knot $K$ correspond to the ones of its mirror image $\tilde{K}$ for a Sp(N)-CS theory:
\beq \label{corspso}
     \langle  W(K) \rangle \Big|_{SO(-N)} = (-1)^{m} \ \langle W (\tilde{K}) \rangle \Big|_{Sp(N)} 
\eeq
For odd-multicomponent knots-links the correspondence hold up to a global sign, where m is the number of components. The mirror image $\tilde{K}$ is needed in order to have opposite the chirality in framing that compensate a sign in the odd terms expansion. In terms of Dubrovnik polynomial (\ref{corspso}) became $D_K|_{SO(-N)}=D_{\tilde{K}}|_{Sp(N)}$, at least for proper knots. So again what suggested by the variational approach can be coherently recovered and extended by the perturbative one.\\ 
The ambient isotopic Dubrovnik-Kauffman polynomial is obtained, as usual, from the regular one thanks to a writhe normalisation: $ a^{-w(K)} D_K$.\\
Another remarkable feature of the variational and perturbative approaches is that allow us to generalise at once the present treatment also to the non-compact groups such as SO(m,n), which are the more interesting ones for describe general relativity in 2+1 dimensions by the Chern-Simons theory. Although from a classical point of view locally isomorphic groups represent the same gauge theory, we have seen as at the quantum level expectation values even of simple knots differ. Thus in case one wants to take advance of the Chern-Simons formalism to study quantum properties of gravity he will have to consider the issue of which is the "good" group election. Actually the values of the fundamental quantities as the Casimirs $c_2 , c_v$, the group's dimension dim$\mathcal{G}$ and the fundamental representation dimension dim($T$) of SO(m,n) are not different from the SO(N) ones, whenever $m+n=N$. Hence the topological quantity $\langle W(K) \rangle$ (\ref{w2ord}) is not affected by the signature change of the Cartan-Killing metric\footnote{Of course a gauge description of gravity needs a further step: also a signature's change in the space-time coordinates, this is more problematic because all the treatment done in this paper is for compact manifolds $\mathcal{M}^3$.}. Up the author knowledge invariant knot polynomials for SO(m,n) groups are not found by means of any other methods; could be interesting to verify it with the help of more rigorous mathematical tools such as quantum groups.
Moreover the SO(m,n) Chern-Simons theory got a richer structure than the SU(N) one. In fact others non-equivalent Chern-Simons Lagrangian can be built from their Chern's characteristic classes apart from the Pontryagin; for instance is possible to use also the Euler or Nieh-Yan topological invariants (see \cite{zanelli} for a review). The expectation values of knotted Wilson loops weighted by this Chern-Simons density remains a topological invariant, but possibly of different kind.

\section{Correspondence with the Potts Model}
\label{correspondence}

\begin{figure}[h]
\begin{center}
\includegraphics[angle=0 , scale=0.7] {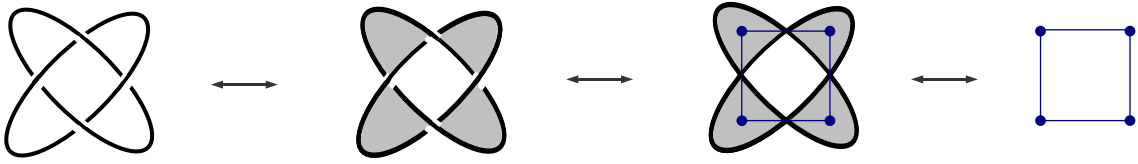}
\end{center}
\caption{\small{K(G) $ \longleftrightarrow $ shading of K(G) $ \longleftrightarrow $ emerging of lattice graph G inside K  $ \longleftrightarrow $ G(K) }}
\label{graph}
\end{figure}

In this section we try to build a bridge between the previous results about first order expectation values of quantum holonomies along a knotted path and some statistical system such as the Potts Model. Of course it is clear that an exact equality can not hold since the Chern-Simons observables are knot invariants while the Potts partition functions are not. Nevertheless something can be said, but at the price of renouncing to the knot topological invariance. First let us remind some fundamental facts about the Potts model that we will be used afterwords. \\
It is found in \cite{kauffman_stat} that the partition function of the Q-Potts Model of a statistical lattice represented by a graph G is the \emph{ Potts state bracket} $\{K(G)\}$ of the knot-link K dual to the graph G. That's because this state bracket expansion coincides exactly with the dichromatic polynomial, or the Tutte polynomial, of the graph G. We remember the definition of the Potts state bracket:
\bea
\label{potbra}
      &i)&    \qquad      \{ \Across \}= Q^{-1/2} v \{ \Asmooth \} + \{ \Bsmooth \} \nn  \\ 
      &ii)&   \qquad      \{ \bigcirc \ K \}=Q^{1/2} \  \{  K \}  \nn \\
      &iii)&  \qquad      \{ \bigcirc \}=Q^{1/2}
\eea
To be more precise for any alternating knot or link K it is possible to construct a graph lattice G(K) checkerboard shading its planar diagram and assigning to each shadow a vertex and for each crossing a bound, as shown in figure \ref{graph}. Vice-versa for any two dimensional graph G one can associate its dual knot K(G). Note that this is a one-to-one\footnote{When the white region is left outside.} mapping between planar graphs and alternate knots and note that any knot got its alternate representative, that is can be drawn as an alternate planar diagram.\\
Thus the Q-Potts partition function for a certain statistical lattice $P_{G}(Q,t)$ is given by the dichromatic polynomial $Z_G(Q,v)$ of its graph G (whenever $v=e^{J/kt}-1$) or by the Potts state bracket of its associated knot $\{K\}$ as follows:
\beq  \label{PottsP}
    P_{G(K)}(Q,t) \ = \ \sum_\sigma e^{\frac{J}{k_B t}\sum_{<i,j>} \delta(\sigma_i,\sigma_j)} \ = \ Q^{V/2} \{K\} (Q,v=e^{J/k_B t}-1)\ ,
\eeq
where $V$ is the number of vertex of the graph (i.e. the number of the lattice's sites or rather the number of shaded region of the knot), $t$ is the temperature, $k_B$ the Boltzmann's constant, $\sigma_n$ is one of the Q possible states of the nth vertex and $J=\pm1$ according to the ferromagnetic or anti-ferromagnetic case.\\

\subsection{SO(-2) \& Sp(2) Holonomies and Q=4 Potts Model}
\label{sp2corr}

First we consider a special case, that is when the Kauffman polynomial reduces to the Kauffman state bracket $[K](q)$ (or to the Jones Polynomial whether writhe normalised), which occurs for the SO(-2), Sp(2) \footnote{Correlated by (\ref{corspso})} or SU(2) Chern-Simons theory, as we have seen in section \ref{varspn} and \ref{perturbative-approach}:
$$ \langle W(K) \rangle (z=q-q^{-1},a=-q^3) = [K](q) \ . $$
Then we perform a shift in the q-variable: $ [K] \rightsquigarrow q^{c(K)} [K]$, where $c(K)$ is the number of crossing in the knot K diagram. This shift is the point where regular isotopical invariance of the Kauffman polynomial is broken. So focusing just on the first order approximation, one gets the following bracket $q^{c(K)} [K](q)\big|_{1^{st}-order}:=\ \ll K \gg (1-i\pi/2k)$:
\bea
\label{<<K>>}
   &i)&  \quad \ll \Across \gg \ = q^2 \ll \Asmooth \gg + \ll \Bsmooth \gg  \ = \Big[1-\frac{i\pi}{k}+O\Big(\frac{1}{k^2}\Big)\Big] \ \ll \Asmooth \gg + \ll \Bsmooth \gg \nn  \\ 
    &ii)& \quad     \ll \bigcirc \ K \gg \ = N +O\Big(\frac{1}{k^2}\Big)\  \ll  K \gg \nn \\
     &iii)& \quad    \ll \bigcirc \gg \ = N +O\Big(\frac{1}{k^2}\Big)
\eea
The analogy with the Potts state bracket (\ref{potbra}) is now evident:
\beq \label{evid}
\{K\}(Q,v)= \ \ll K \gg (\pm v^{1/2} Q^{-1/4})  .
\eeq
Let now concentrate on the SO(-2) case, such that once the q-shift is reabsorbed one recovers knot invariance, so $Q=N^2=4$.
Using (\ref{PottsP}) and (\ref{evid}) it is easy to see that $ -2^V \ll K \gg$ represents the Q=4 Potts partition function for the lattice graph associated to the knot K. In terms of the first order Wilson loops expansion it reads:
\beq   \label{sp2pot}    P_{G(K)} = Q^{V/2}  \{ K \}  = N^V  \ q^{c(K)} \  \langle W(K) \rangle \Big|_{1^{st} - \mathrm{order}} \eeq

An example may get things clearer: consider a 2x2 lattice graph G of figure \ref{graph} and its dual knot-link $K(G)$ (with $V=4$). From skein relations (\ref{potbra}) (or equally from the deletion-contraction rule that define the dichromatic polynomial $Z_G(4,v)$) one gets the Q=4 Potts partition function for the graph $G(K)$: 
\beq  \label{zg4v}
      Z_G(4,v) = 4^{V/2} \{K\}= 4^2 (4^2 + 4 \cdot 4 v + 6 v^2+ 4 \cdot 4^{-1} v^3 + 4^{-1} v^4)
\eeq
while from the skein relations (\ref{<<K>>}) one get the expectation value of the holonomy along the knot $K(G)$, up to $O(1/k^2)$:
\beq 
    -2^V  \ q^{c(K)} \  \langle W(K) \rangle \Big|_{1^{st} - \mathrm{ord}} = 2^4 \big(1-\frac{i2\pi}{k}\big) \Big[16 \big(1+\frac{i2\pi}{k}\big) - 32 \big(1+\frac{i\pi}{k}\big) + 24 - 8 \big(1-\frac{i \pi}{k}\big) + 4 \big(1-\frac{i 2 \pi}{k}\big) \Big]  \nn 
\eeq
It's easy to see that (\ref{sp2pot}) is fullfilled imposing $v= -2+i2\pi/k$ in (\ref{zg4v}). So the first order expectation value of the Wilson loop along a knotted path K for a SO(-2)/Sp(2) Chern Simons theory can be extracted from the partition function of a Q=4 Potts model of a lattice graph $G(K)$ dual to the knot $K$, and vice-versa. This correspondence works well for any two dimensional lattice graph, not just for regular ones like the sample presented in figure \ref{graph}.\\
Even thought $\langle W(K) \rangle \big|_{1^{st} - \mathrm{order}} $  and $P_G(K)$ are not exactly the same they share some features, for instance their zeroes. So $\langle W(K) \rangle \big|_{1^{st}} $'s zeros can be interpreted as the Fisher zeros of the statistical lattice associated to $K$, which encode many important physical properties of the system. Also the critical temperature $t_c$ (when the statistical system acquires conformal invariance) of the Potts model can be easily read: In the knot formalism it occurs where $\langle W(K) \rangle=\langle W(\tilde{K}) \rangle$, that is when $1-i\pi/k=1$, so in the limit $k \rightarrow \infty$, which means $t_c=\frac{J}{k_B}\frac{1}{\textrm{ln}(\sqrt{Q}+1)}$.\\
It's worth remark at this point that the SO(-2)/Sp(2) group (or even SU(2)) gives rise to the Jones polynomial too. This polynomial (at the non-perturbative level) is known to describe the partition function of a particular kind of Potts model with two Boltzmann factor, which is of different kind respect to the standard Potts model considered here (see \cite{wu-knot} and \cite{wu-potts}).\\
The correspondence holds also at the following orders of the perturbative expansion, basically in the same way it works at the first order. For instance one can obtain $\langle W(K) \rangle \big|_{2^{sd}-order}$ from the Q=4 Potts partition function identifying $v$ and $Q$ as follows:
\bea
     v   &\leftrightsquigarrow& - 2 \left[ 1 - \frac{i\pi}{k} - \left( \frac{\pi}{k} \right)^2 + O \left( \frac{1}{k^3} \right)  \right] \nn \\
            Q^{\frac{1}{2}} &\leftrightsquigarrow& - 2 \left[ 1 - \frac{1}{2} \left( \frac{\pi}{k} \right)^2 + O \left( \frac{1}{k^3} \right)  \right]  \nn 
\eea 
The simple relation between $Q$ and $N$ is now spoiled and moreover this fact makes the analogy between the two models purely formal because choosing a particular Q imply fixing at the same time the temperature to a constant value.

\subsection{Sp(N) holonomies and Q-Potts Model}

We would like to do something similar to previous subsection, but for generic $N$. Now that procedure is less direct because the Kauffman polynomial can not be cast in a simple form such as the state bracket [K]. To connect the two theories, in particular to give the Q-Potts partition function a similar structure to the Dubrovnik polynomial one,  we can introduce a new bracket polynomial $\| K \|(Q,v)$ defined by the following skein relations:
\bea
\label{|K|}
          &i)&    \quad    \|\Across \| - \| \Bcross \| \ = (Q^{-1/4}v^{1/2}-Q^{1/4}v^{-1/2}) \big[ \| \Asmooth \|  - \| \Bsmooth \| \big] \nn  \\ 
          &ii)&   \quad    \| \Rcurl \|  =   (Q^{1/4}v^{1/2}+Q^{1/4}v^{-1/2}) \| \Arc\|  \ \ , \quad \| \Lcurl \|= (Q^{-1/4}v^{1/2}+Q^{3/4}v^{-1/2})  \|\Arc \|\nn \\
          &iii)&  \quad    \| \bigcirc \| = Q^{1/2}  \nn  \\
          &iv)&   \quad    \| \duecross \| = (Q^{-1/2}v+Q^{1/2}+Q^{1/2}v^{-1}) \ \| \Asmooth \| +   \ \| \Bsmooth \| 
\eea
The Q-Potts partition function, in character of the dichromatic polynomial $Z_{G(K)}(Q,v)$, has the following form in term of $\| K \|$:
$$ Z_{G}(Q,T) = Q^{V/2} [Q^{-1/4}v^{1/2}]^{c(K)} \| K \| .$$
Even in this form $\|K\|$ is not a isotopical invariant of the knots, as $\langle W(K) \rangle$ because the two coefficients in (\ref{|K|}-$ii$) are not reciprocal and (\ref{|K|}-$iv$) does not satisfy the second Reidemeister move. However there is a point where both (\ref{|K|}-$ii,iv$) becomes invariant, that is for $v=(-Q\pm\sqrt{Q^2-4Q})/2$. This value of the temperature is exactly the one that relates the Potts model to the Khovanov homology \cite{kauff_khovanov}. Comparing the $\|K\|(Q,v)$ bracket with the first order expectation value of the holonomy $\langle W(K) \rangle \big|_{1^{st-ord}}$ one has to impose $Q=N^2$ and $v=N(1-i\pi/k)$. So the $\| K \|(N,k)$ invariance occurs, in terms of the Chern-Simons coupling constant $k$ and the fundamental representation dimension $N$, just for $N=-2$, i.e the previous case we analysed in section \ref{sp2corr}. \\
Therefore for a generic $Q=N^2\neq4$ is not possible to pass from the Potts partition function to the first order Wilson loop expectation value as we did for the $SO(N)/Sp(2)$ case. What can be done at most is define a generic bracket polynomial which include both $P_G$ and $\langle W(K) \rangle$ and specialises to one or the another for some values of its variables. This is done in appendix \ref{appM}.

\section{Comments and Conclusions}
In this paper is analysed the relation between expectation values of Wilson loop in three-dimensional SO(N) Chern-Simons field theory and an isotopic invariant of knots, the Kauffman polynomial. This equivalence is achieved in a simple intuitive knot variational approach borrowed by \cite{cotta}'s and \cite{kauffman}'s scheme which was elaborated for obtaining the Witten result: HOMFLY polynomial from the SU(N) gauge group. The key point of this construction is based on the existence of a Fierz identity for the infinitesimal generators of the group in certain representations. With precisely the same interpretation of the expectation value's path variations and no other extra assumptions respect to the original work, here we exactly get the conformal field theory known result for SO(N): Kauffman polynomial. 
It suggests that the easy variational knot approach, expressly built for SU(N), works well also for different gauge group theories as SO(N). So its heuristic geometrical assumptions are endorsed.\\
Convinced of all that and encouraged by negative dimension group theory suggestion we explored also the Sp(N) group getting the exact skein relation. In particular in the simple Sp(2) case we are able to find its isotopic invariant: the original Jones Polynomial.\\
Furthermore to enforce and extend those results, an independent procedure has been performed, the quantum field theory method can not only full recover  the variational approach but also: improve its outcomes precision of an order of magnitude, extend to groups with semi-definite Cartan-Killing metric  as well Sp(N) with $N \neq 2$ and most of all prove, up to $O(1/k^3)$, the correspondence between isotopy invariant polynomials from SO(N) and Sp(-N) Chern-Simons theories.\\
To sum up, these procedures give for SU(N), SO(N)/Sp(N) and Sp(2) the famous HOMFLY, Kauffman and Jones polynomials respectively. Hence they may be used for other groups or representations to find new link invariants, both based on skein relations or not. This could give new insights into knots theory, which is still looking for a link invariant able to distinguish conclusively knots isotopic equivalence. \\
From a physical point of view it's interesting to note that not only the Jones polynomial, at non perturbative level, correspond to the partition function of the Potts model with two Boltzmann weight factors, but also its first order perturbation expansion, in the realm of the Chern-Simons theory, gives the standard Q=4 Potts partition function (and vice-versa).
Moreover the connection between the quantum holonomies of Sp(2) Chern-Simons theory and the $Q=4$ Potts partition function opens the possibility to relate apparently disconnected physical systems. This is actually the main motivation of the author.  In fact, since \cite{AcTo}, it is well known that Sp(2)$\times$Sp(2) Chern-Simons theory describes 2+1 gravity with negative cosmological constant. Furthermore the first terms in the Kauffman bracket expansion give states of 3+1 quantum gravity in the loop representation \cite{ga-pu}. This feature of knot theory may represent the tip of an iceberg that links discrete statistical models with the expectation value of holonomies of gravitational theories. Work in this direction is in progress.\\

\section*{Acknowledgements}
\small
I would like to thank Louis Kauffman, Roberto Troncoso, Steven Willison and Jorge Zanelli for fruitful discussions.\\
The Centro de Estudios Cient\'{\i}ficos (CECS) is funded by the Chilean Government through the Millennium Science Initiative and the Centers of Excellence Base Financing Program of Conicyt and by Conicyt grant "Southern Theoretical Physics Laboratory" ACT-91.
CECS is also supported by a group of private companies which at present includes Antofagasta Minerals, Arauco, Empresas CMPC, Indura, Naviera Ultragas and Telef\'{o}nica del Sur.
\normalsize

\begin{appendix}

\section{General Potts-Dubrovnik polynomial $M_K$}
\label{appM}

Define the following bracket polynomial $M_K(a,b,c,d,z)$:
\bea
\label{M}
          &i)&    \quad    M( \Across ) - M(\Bcross) \ = z \big[ M(\Asmooth)  - M(\Bsmooth) \big] \nn  \\ 
          &ii)&   \quad    M(\Rcurl)  = a \ M(\Arc) \quad , \quad M(\Lcurl)= b \ M(\Arc)\nn \\
          &iii)&  \quad    M(\bigcirc) = d    \\
          &iv)&   \quad    M(\duecross) = c \ M(\Asmooth) + \ M(\Bsmooth) \nn
\eea
$M_K$ reduces to the Kauffman-Dubrovnik polynomial when $b=a^{-1} , \ c=0  , \ d=1$; while to $\langle W(K) \rangle$ when $d=(a-a^{-1})/z + 1$. So for those values of the variables it is an invariant of regular isotopy. But the Potts partition function is not invariant so this latter has $b\neq a^{-1}$ and $c$ switched on, as can see in the following table, where two different specialisations of the $M_K$ polynomial are shown: 

\vspace{5mm}
\hspace{-0.5cm}
\label{tableM}
\begin{tabular}{|c|ccccc|}
	\hline
 $M_K$ &  $a$  &  $b$  &  $c$  & \hspace{-1cm} $d$ &\hspace{-1cm} $z$ \\
	\hline
$\langle W(K)\rangle$   & $\alpha$ & $\alpha^{-1}$ & $0$ & \hspace{-1cm} $(a-a^{-1})/z + 1$ & \hspace{-10mm} $-i\pi/k$   \\
$\|K(G)\|$                 & $Q^{\frac{1}{4}}v^{\frac{1}{2}}+Q^{\frac{1}{4}}v^{\frac{-1}{2}}$ & $Q^{\frac{-1}{4}}v^{\frac{1}{2}}+Q^{\frac{3}{4}}v^{\frac{-1}{2}}$ & $Q^{-\frac{1}{2}}v+Q^{\frac{1}{2}}+Q^{\frac{1}{2}}v^{-1}$ & \hspace{-10mm} $Q^{\frac{1}{2}}$ & \hspace{-7mm} $Q^{-\frac{1}{4}}v^{\frac{1}{2}}-Q^{\frac{1}{4}}v^{-\frac{1}{2}}$      \\
	\hline
\end{tabular}

\vspace{5mm}

\end{appendix}


\end{document}